\documentclass[12pt]{article}

\advance\textwidth by 0.5in
\advance\hoffset by -0.25in

\usepackage{epsf}

\newcommand{\be}{\begin{eqnarray}}
\newcommand{\ee}{\end{eqnarray}}

\newcommand{\slpartial}{\raise.15ex\hbox{$/$}\kern-.53em\hbox{$\partial$}}
\newcommand{\slA}{\raise.15ex\hbox{$/$}\kern-.73em\hbox{$A$}}
\newcommand{\slp}{\raise.15ex\hbox{$/$}\kern-.57em\hbox{$p$}_1}

\begin{document}

\thispagestyle{empty}
\title {\bf Forward Quark Jets from Protons Shattering the Colored Glass}

\author
{
Adrian Dumitru and Jamal Jalilian-Marian
 \\
 {\small\it Physics Department, Brookhaven National Lab,
            Upton, NY 11973, USA}\\
}

 \maketitle

\begin{abstract}

\noindent 
We consider the single-inclusive minijet cross section in $pA$
at forward rapidity within the Color
Glass Condensate model of high energy collisions. We show that
the nucleus appears black to the incident quarks except for very
large impact parameters. A markedly flatter $p_t$ distribution as
compared to QCD in the dilute perturbative limit is predicted for
transverse momenta about the saturation scale, which could be as large as
$Q_s^2 \simeq 10$~GeV$^2$ for a gold nucleus boosted to rapidity $\sim 10$
(as at the BNL-RHIC).
\end{abstract}

QCD correctly predicted logarithmic violations of Bjorken scaling
in Deep-Inelastic electron-proton scattering at very large $Q^2$, i.e.\
at very short distances~\cite{gw}. Asymptotic freedom provides the 
theoretical basis for the successfull applications of perturbative  
QCD to hard scattering, short distance phenomena. However, 
the region of QCD phase space where the field strengths are strong
is largely unexplored. This is where one expects that
cross sections become comparable to geometric sizes of hadrons and nuclei
(the ``black limit'') and where the unitarity limit is reached. 
A perturbative QCD based mechanism for unitarization of cross sections is 
provided by gluon saturation effects~\cite{glr,mq}. A semi-classical approach
to gluon saturation and QCD in the high energy limit (small $x$) 
was developed in~\cite{mv,yk,jk,ianc} and applied to high energy heavy 
ion collisions at RHIC \cite{kmw,kv,kl,dm}.    

An ideal environment to study gluon saturation and unitarization 
effects is provided by large nuclei since high gluon density effects
are expected to be amplified by a factor of $A^{1/3}$ due to the Lorentz
contraction of the nucleus. The scale associated
with the high gluon density, the saturation scale $Q_s$, grows
with energy and $A$ and decreases with increasing impact parameter.
The Relativistic Heavy-Ion Collider (RHIC) at BNL will
soon allow experimental study of proton-gold or deuteron-gold
collisions at a center-of-mass energy of $\sqrt{s} \sim 200-300$~GeV. We 
suggest that the saturation regime of QCD can be probed at RHIC by 
measuring the inclusive cross section in $p+Au$ (or $d+Au$) collisions
(in this regard, see also~\cite{dm,pA_2}).
In particular,
in the forward region, i.e.\ close to the rapidity of the proton beam,
the saturation scale $Q_s$ can become quite large due to renormalization group
evolution in rapidity~\cite{jk,ianc}.
Thus, we predict significant modifications of the
$p_t$ distribution of produced pions relative to leading twist
perturbation theory at transverse momenta as large as several GeV.
A modification of the {\em longitudinal} distribution of leading
hadrons produced by electrons scattering inelastically from a black target
has been
predicted previously~\cite{fgs}. Here, we focus on the transverse distribution
in the forward region from $p+A$ scattering, which will be analyzed
experimentally in the near future at RHIC.

At large rapidity, we consider the quark-nucleus elastic and total 
scattering cross sections. (In turn, towards midrapidity gluon production
becomes the dominant contribution to the cross section in the Color Glass
Condensate model~\cite{dm}.)
We argue that the total quark-nucleus scattering cross
section may be related to the single inclusive hadron (jet) cross section
in proton-nucleus collisions by using the colinear factorization theorem 
on the proton side. Let $p^\mu$ ($q^\mu$) be the momentum of the
incoming (outgoing) quark. We assume the quark is moving along
the left branch of the light cone such that $p^-\gg p^+=p_t^2/2p^-$.
The starting point is the scattering amplitude (for brevity, we do not write
polarization indices explicitly)
\be
\langle q(q)_{out}|q(p)_{in}\rangle =
\langle out|b_{out}(q)b^{\dagger}_{in}(p)|in\rangle~,
\label{eq:amp}
\ee
which, using the LSZ formalism \cite{iz} can be written as
\be
\langle out|b_{out}(q)b^{\dagger}_{in}(p)|in\rangle &=&
- { 1 \over Z_2} \int d^4x d^4y  e^{-i(p x - q y)}
\bar{u}(q)[i \stackrel{\rightarrow}{\slpartial}_y -m]
\nonumber \\ 
&&\langle out|T \psi (y) \bar{\psi}(x)|in\rangle  
[-i \stackrel{\leftarrow}{\slpartial}_x -m] u(p)
\label{eq:general}
\ee
where $m$ is the quark mass and $Z_2$ is the fermion wave function
renormalization factor. $u(p)$ is the quark spinor with momentum $p$.
The fermion propagator $G_F$ in the background of the classical color 
field is 
\be
\langle out|T \psi (y) \bar{\psi}(x) |in\rangle  \equiv -i \langle  out |
 in\rangle  G_F (y,x)~.
\ee
The amplitude then becomes
\be
\langle q(q)_{out}|q(p)_{in}\rangle &=&
{i \over Z_2}  \langle  out | in\rangle  \int d^4x d^4y  e^{-i(p x - q y)} 
\nonumber \\
&&\bar{u}(q)[i \stackrel{\rightarrow}{\slpartial}_y -m] 
G_F(y,x) [-i \stackrel{\leftarrow}{\slpartial}_x -m] u(p)~.
\label{eq:amplitude}
\ee
In momentum space, the fermion propagator $G_F$ can be written
as~\cite{mv,hw,gp}
\be
G_F(q,p)= (2\pi)^4 \delta^4 (q-p) G^0_F (p) -ig G^0_F (q)
\int {d^4k \over (2\pi)^4} {\slA}(k) G_F (q+k,p)~,
\label{eq:Geqmom}   
\ee
where ${\slA}=A^{\mu} \gamma_{\mu}$ is the classical background 
color field, and $G^0_F$ is the free fermion propagator. 
It is useful to define the interaction part of the fermion propagator 
from (\ref{eq:Geqmom}) as
\be
G_F(q,p)= (2\pi)^4 \delta^4 (q-p) G^0_F (p) + G^0_F (q)
\tau (q,p) G^0_F (p)~.
\label{eq:Gint}
\ee
Substituting (\ref{eq:Gint}) into the amplitude (\ref{eq:amplitude})
leads to
\be
\langle q(q)_{out}|q(p)_{in}\rangle  =  \bar{u}(q) \tau (q,p) u(p)~.
\label{eq:finalamp}
\ee
Here, we have set $Z_2 =1$ and $ \langle out | in\rangle =1$ since
we are working to leading order in $\alpha_s$ and our background
field is time independent.
This is a very simple relation between the amplitude  
for scattering of a quark or anti-quark from the Color Glass
Condensate and the quark propagator in the background color
field of the nucleus.

The explicit form of the quark propagator in the background
of a classical color field was
calculated in \cite{mv,hw,gp}. The interaction part,
as defined in (\ref{eq:Gint}) is given by \cite{gp}
\be
\tau (q,p)=(2\pi) \delta (p^- - q^-) \gamma^- 
\int d^2 z_t \bigg [ V (z_t) -1 \bigg ]
e^{i(q_{t} - p_{t}) z_t}~,
\label{eq:taures}
\ee
where 
\be
V(z_t) \equiv \hat{P} \exp \bigg [-ig^2 \int^{+\infty}_{-\infty}
d z^- {1 \over \partial^2_t} \rho_a (z^-,z_t) t_a \bigg]~,
\label{eq:udef}
\ee
and $t_a$ are in the fundamental representation. Using (\ref{eq:taures})
in the scattering amplitude (\ref{eq:finalamp}) gives
\be
\langle q(q)_{out}|q(p)_{in}\rangle  =
(2\pi) \delta (p^- - q^-)
\bar{u}(q) \gamma^-   u(p) \int d^2 z_t 
\bigg [ V(z_t) -1 \bigg ] e^{i(q_{t} - p_{t}) z_t}~.
\label{eq:expamp}
\ee
The presence of the delta function in the amplitude is due to 
the target being (light-cone) time independent which leads
to conservation of the ``minus'' component of momenta. 
It can be factored out in the standard fashion,
\be
\langle q(q)_{out}|q(p)_{in}\rangle  &=& (2\pi) \delta (p^- - q^-) M (p,q)~,
\label{eq:mamp}
\ee
which gives the cross section
\be
d\sigma = \int {d^4 q \over (2\pi)^4} (2\pi) \delta (2 q^+ q^- -q_t^2)
\theta (q^+) {1 \over 2p^-} (2\pi) \delta (p^- - q^-) 
|M (p,q)|^2~.
\label{eq:diffcs}
\ee
The local density of color charge in the nucleus, $\rho_a(x_t,x^-)$, is
a stochastic variable which has to be averaged over. Commonly, one assumes
a distribution of the charge sources which is local and
Gaussian~\cite{mv,dm,gp,km,kw},
such that the average of any operator $O$ is
\be
\langle O\rangle = \int {\cal D}\rho~O[\rho] 
\exp\left(- {{\rm tr}\, \rho^2 / \mu^2}\right)~.
\ee
$\mu^2(z_t,x^-)$ denotes the color charge density per unit transverse area
$d^2 z_t$, and per unit of rapidity, $dx^-/x^-$, in the nucleus.
When computing the total cross section, we will have to square the 
amplitude~(\ref{eq:expamp}) before averaging over the color charge
density $\rho$ of the classical background field. On the other hand,
for the case of elastic scattering, we first average the
amplitude~(\ref{eq:expamp}) over $\rho$ and then square it~\cite{km,bmh}. 
That way no color exchange occurs over a large distance in rapidity (from the
projectile quark to the nucleus).

The averages of $V(z_t)$ and $V^{\dagger}(z_t) V(\bar{z_t})$
are given by~\cite{gp,kw}
\be
\langle  V(z_t) \rangle _{\rho} = \exp \bigg [ - {g^4 (N_c^2 -1) \over 4 N_c}
\chi \int d^2 y_t \, G^2_0 (z_t - y_t)\bigg]~,
\label{eq:Vave}
\ee
and
\be
\langle  V^{\dagger}(z_t) V(\bar{z_t}) \rangle _{\rho} =  
\exp \bigg [ - {g^4 (N_c^2 -1) \over 4 N_c}
\chi  \int d^2 y_t \,
[G_0 (z_t - y_t) - G_0 (\bar{z_t} - y_t)]^2 \bigg]~.
\label{eq:VVave}
\ee
We have defined $\chi(x^-)
\equiv \int_{x^-/x_0^-}^{x_A^-/x_0^-} d z^- \mu^2(z^-) $, which
is the density of color charge in the nucleus per unit transverse area
{\em integrated} over longitudinal phase space (rapidity).
$x_A^-\ll x_0^-$ is the coordinate of the nucleus, with $y_A=\log x_0^-/x_A^-$
its rapidity; while $x^-\ll x_0^-$ is the coordinate of the quark projectile,
and $y=\log x^-/x_0^-$ is its rapidity. ($x_0^-$ is a reference point, which we
choose to be midrapidity). 
$G_0 (z_t - y_t)$ is the free propagator of static gluons 
\be
G_0 (z_t - y_t) = - \int\limits_{\Lambda_{QCD}^2} {d^2 k_t \over (2\pi)^2}
{e^{ik_t(z_t - y_t)} \over k_t^2}~.
\label{eq:freeG}
\ee 
The trace of the quark spinors in the squared amplitude is 
\be
\frac{1}{2}
\sum\limits_{spins}\left| \bar{u}(q) \gamma^-  u(p) \right|^2
 = 4 p^- q^-~.
\label{spinsum}
\ee
Averaging over the colors of the incoming quark is made trivial by the fact
that~(\ref{eq:Vave},\ref{eq:VVave}) are diagonal in color space.
For elastic scattering, we shall use~(\ref{eq:Vave}) to
color average the amplitude given by~(\ref{eq:expamp}), and afterwards
square it. The color averaging of the amplitude leads to a delta-function of
transverse momenta $(2\pi)^2\delta^2 (p_t -q_t)$. This is due to the assumed
translational
invariance of the target in transverse space (we have assumed a large
cylindrical target nucleus so that the color charge density is uniform)
and should be understood as $\pi R_A^2 (2\pi)^2 \delta^2 (p_t - q_t)$ 
in the squared amplitude. Using (\ref{eq:expamp}),  (\ref{eq:mamp}) and 
(\ref{eq:diffcs}) finally leads to 
\be
{d\sigma^{el}_{qA} \over d^2b} =  
\bigg[1 -  e^{- \pi^2 Q_s^2/N_c\Lambda_{QCD}^2} \bigg]^2
\label{eq:sigdiff}
\ee
where we have introduced the saturation scale of the target,
$Q_s^2\equiv  (N_c^2 -1) \alpha_s^2 \chi/\pi$ (this is the same definition as
in~\cite{kl}).

To compute the total cross section for scattering of the quark on the
nucleus, we first square the amplitude (\ref{eq:expamp})
and then average over the background field
using~(\ref{eq:Vave},\ref{eq:VVave}). It leads to  
\be
{d\sigma^{tot}_{qA} \over d^2b} =  
\int\!\! {d^2q_{t} \over (2\pi)^2} 
\int\!\! d^2 r_t \,
e^{-iq_{t} r_t} 
\bigg[ e^{- {2 \pi Q_s^2 \over N_c} \int 
{d^2k_t / k_t^4}~\big[1 - \exp(ik_t r_t)\big]} - 
2 e^{- \pi^2 Q_s^2/N_c\Lambda_{QCD}^2} + 1 \bigg]~.
\label{eq:siginc}
\ee
The integral over $q_t$ just gives $\delta^2 (r_t)$
which in turn can be used to perform the $r_t$ integration. The final result
for the total cross section is
\be
{d\sigma^{tot}_{qA} \over d^2b} =  
2 \bigg[1 -  e^{- \pi^2 Q_s^2/N_c\Lambda_{QCD}^2} \bigg] .
\label{eq:finalsiginc}
\ee
{}From eqs.~(\ref{eq:sigdiff}) and (\ref{eq:finalsiginc}), we see that in the
high energy limit ($Q_s \rightarrow \infty$) we have 
\be
{\sigma_{qA}^{tot} } = 2 {\sigma_{qA}^{el}}= 2\pi R_{A}^2 
\ee
as a consequence of unitarity. 

It is more interesting to consider the {\it differential} cross section 
${d\sigma_{qA}^{tot}/ d^2b d^2 q_t}$ when the quark momentum $q_t$
is large ($q_t^2 \gg \Lambda_{QCD}^2$). In this limit, one can convolute
the quark-nucleus cross section with the quark distribution function
in a proton and the quark fragmentation function into hadrons, thereby
relating $qA$ scattering to single inclusive hadron (jet) production in $pA$ 
collisions \cite{jq}.

At very large transverse momentum, $q_t^2 \gg Q^2_s$,
we can expand the $V$'s in eq.~(\ref{eq:VVave}), keeping only the
first non-trivial term. This corresponds to the dilute perturbative
limit. Using~(\ref{eq:freeG}) then gives the expected
\be
{d\sigma_{qA} \over d^2 q_t d^2b} \sim  {Q_s^2 \over q_t^4}~.
\label{pQCD}
\ee
In the region $ q_t^2 \sim Q^2_s\gg \Lambda_{QCD}^2$, in turn, we must resum
higher twists, i.e.\ rescatterings in the nuclear field. We then obtain
\be
{d\sigma_{qA} \over d^2 q_t d^2b} \sim  {1 \over q_t^2}~.
\label{qA_sat}
\ee
Clearly, the non-linearities of the classical field have flattened the
differential cross section as compared to the perturbative
cross section~(\ref{pQCD}). This arises, in fact, from a {\em suppression} of
particle production. It should be easy to distinguish from
$p_t$-broadening, i.e.\
``initial-state'' interactions of the beam quarks with spectators from the
target~\cite{bbl}, which {\em enhance} the cross section at semi-high
$p_t$.

The single inclusive $p+A\rightarrow h +X$
cross section can now be obtained in principle by convoluting the total $qA$
cross section with the quark 
distribution function of the proton at the factorization scale $Q_f^2$,
which can for example be chosen to be $q_t^2$,
and the quark-hadron (jet) fragmentation function $D_{q/h}(z,Q_f^2)$,
where $z$ is the ratio of hadron and quark momenta.
\be \label{pA_SD}
{d\sigma^{pA\rightarrow h X} \over dy d^2 k_t d^2b} =
\int dx {dz \over z^2} \; q(x,Q_f^2) 
\; {d\sigma^{tot}_{qA} \over dy_q d^2 q_t d^2b}
\; D_{q/h}(z,Q_f^2) ~.
\ee
Here, $x$ denotes the fractional (light-cone) momentum carried by the quark,
such that the light-cone coordinate of the incident proton is related to that
of its quark by $x^-_p = x x^-$.

The $qA$ cross sections depend on $x$ through
the saturation momentum of the nucleus.
{}From the analysis of HERA DIS data, the saturation momentum scales like 
$Q_s^2(x^-/x_0^-)/Q_s^2(x_0^-)=(x_0^-/x^-)^\lambda = (x_0^- x/x_p^-)^\lambda$,
with $\lambda\simeq0.3$~\cite{GBW}.
For gold nuclei, it has been estimated that $Q_s^2(x_0^-)\approx2$~GeV$^2$ at
RHIC energy (100A~GeV gold beam, corresponding to $y_A=5.4$)~\cite{kl}.
Therefore, near the rapidity of the incident proton, i.e.\ $y_p=
\log x_p^-/x_0^- =-5.4$, we estimate
that roughly $Q_s^2\approx 10$~GeV$^2$. With $1/\Lambda_{QCD}^2\simeq
25$~GeV$^{-2}$, the exponent in eqs.~(\ref{eq:sigdiff},\ref{eq:finalsiginc})
is $<-800$.
In fact, even at $b\simeq 6$~fm from the center of a gold nucleus,
using a Gaussian density distribution for the nucleus we estimate
that $Q_s^2(x_0^-,b\sim6$~fm$)\simeq 0.75$~GeV$^2$, and therefore $Q_s^2(x_p^-,
b\sim6$~fm$)\simeq 3.8$~GeV$^2\simeq100\Lambda_{QCD}^2$. At the edge of the
acceptance of the BRAHMS spectrometer at RHIC, $y=\log x_0^-/x^-=4$, one
obtains $Q_s^2(x^-,b\sim0)\simeq 6.6$~GeV$^2$.
That is, $2\sigma^{el}_{qA}=\sigma^{tot}_{qA}
=2\pi R_A^2$, with $R_A$ the radius of the
black disc formed by the nucleus (i.e.\ the distance from the center of the
nucleus where $Q_s^2$ becomes of order $\Lambda_{QCD}^2$), is quite close
to the geometric cross section of the gold nucleus.

The quark distribution functions in the proton, and their fragmentation
into pions will modify the transverse momentum dependence of the pion
cross section~(\ref{pA_SD}) relative to the quark cross section~(\ref{qA_sat}).
Nevertheless, that modification is the same for both the dilute perturbative
estimate of the $qA$ cross section~(\ref{pQCD}) as well as for the resummed
$qA$ cross section from~(\ref{qA_sat}). Therefore, a modification of the
$p_t$ distribution from strong nonlinear color fields as compared to
the dilute limit should hold independently of any scale dependence of the
quark distribution and fragmentation functions. The big advantage of
measurements in the forward region is that these effects extend much farther
in $p_t$ than in the central rapidity region, hopefully making it much less
ambiguous to observe them experimentally.
\\[1cm]
{\bf Acknowledgement:}\\
We thank F.~Gelis, S.~Jeon, D.~Kharzeev,
Y.~Kovchegov, A.~Kovner, 
L.~McLerran, J.~Qiu and R.~Venugopalan for useful discussions. 
This work is supported by the U.S.\ Department of Energy under Contract 
No.\ DE-AC02-98CH10886. J.J-M.\ is also supported in part by a LDRD and a 
PDF from BSA.

\end{document}